\documentstyle[12pt]{article}

\def\beqa{\begin{eqnarray}}
\def\eqa{\end{eqnarray}}
\newcommand{\kb}{\mbox{$\underline {k}$}}
\newcommand{\xb}{\mbox{$\underline {x}$}}

\newcommand{\de}{\mbox{$\frac{1}{2}$}}

\def\beq{\begin{equation}}
\def\eq{\end{equation}}

\begin{document}
\input epsf

\begin{titlepage}
\vspace*{-1.5cm}

\begin{center}
\baselineskip=13pt

{\hspace*{13cm} Saclay-T96/048}
\vspace{2cm}

{\Large \bf  QCD dipole predictions for quark singlet, gluon and $F_L/F_T$
 	distributions at HERA.\\}
\vskip2.5cm
{\Large Samuel Wallon}\\
\vskip1.5cm
{\it Service de Physique Th\'eorique\footnote{Laboratoire de la Direction
des
Sciences de la Mati\`ere du Commissariat \`a l'Energie
Atomique}, CE-Saclay \\ 91191 Gif-sur-Yvette, France}\\
\end{center}
\vspace*{4.5cm}

\begin{abstract} 
In this contribution we apply the QCD dipole picture combined with $k_T$-factorization to get predictions for deep-inelastic scattering on
an onium. Assuming renormalization-group factorization, we get predictions for the $F_2, F_G$ and $R= F_L/F_T$ proton structure functions. We obtain a three-parameter fit of the 1994 H1 data in the low-$x_{bj}$, moderate-$Q^2$ range. $F_G/F_2$ and $R$ are then predicted without any additionnal parameter. The BFKL
dynamics contained in the dipole model is shown to provide a relevant model in describing
the HERA data. The prediction for $F_2$ and $F_G$ are compatible with next-to-leading order DGLAP analysis. By contrast, $R$ is expected to be much
lower at small $x_{bj}$. 

\vskip3.5cm
\noindent
Talk given at the XXXIth Rencontres de Moriond

\noindent
QCD and High Energy Hadronic Interactions 

\noindent
March 23-30, 1996, Les Arcs, France

\end{abstract}

\end{titlepage}

\renewenvironment{thebibliography}[1]
        {\begin{list}{\arabic{enumi}.}
        {\usecounter{enumi}\setlength{\parsep}{0pt}
\setlength{\leftmargin .75cm}{\rightmargin 0pt}
         \setlength{\itemsep}{0pt} \settowidth
        {\labelwidth}{#1.}\sloppy}}{\end{list}}

\newpage
\setcounter{equation}{0}

\section{Introduction}
In this contribution, we show that the QCD dipole picture \cite{mueller1,mueller2,mueller3,nik}, which contains the BFKL dynamics, is a pertinent model for describing the proton structure functions at HERA in the low-$x_{bj}$ and moderate-$Q^2$ range. The most recent HERA 1994 published data \cite {h1} provide a possibility of distinguishing
between the different QCD models for small-$x_{bj}$ physics. Besides the DGLAP evolution
equations \cite{dglap} based on the renormalization group evolution, the BFKL hard Pomeron \cite{lip} should be relevant at small $x_{bj}$. It is thus a challenge to get observables
which could distinguish between these two different theoretical predictions,
and to compare these observables to the data. The perturbative dipole model, combined with the $k_T$-factorization (or Regge factorization) \cite{catani}, allows
us to implement the BFKL dynamics into deep inelastic scattering. We then get predictions for $F_2, F_G$ and $R = F_L/F_T$ proton structure functions. 

\section{BFKL dynamics in the QCD dipole framework}

\begin{figure}[htb]
\centering
\epsfysize=12.0cm{\centerline{\epsfbox{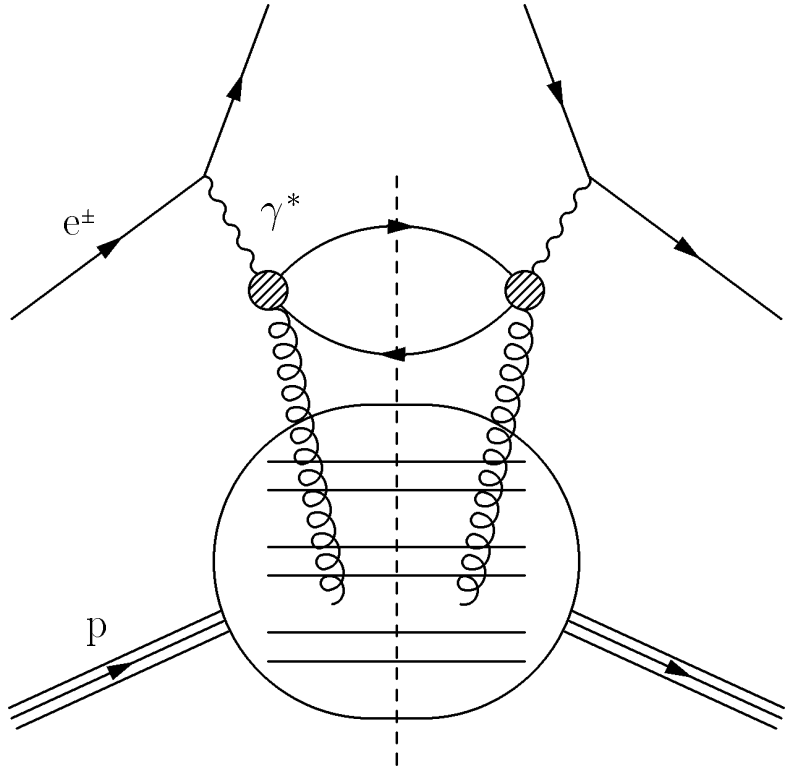}}}
\caption{$k_T$-factorization and dipole model applied to $e^{\pm}-p$ deep inelastic scattering.}
\label{graphef2}
\end{figure}

In this section, we perform an analyse of $e^\pm - p$ deep inelastic scattering at low $x_{bj}$,
based on $k_T$-factorization and dipole color model, as illustrated in figure \ref{graphef2}.

\noindent
In the Regge limit, one can apply the $k_T$-factorization tool \cite{catani} in order to
extract a photon of virtuality $Q^2$ off a proton. It involves the elementary 
Born  cross-section $\hat{\sigma}_{\gamma g}/Q^2$ of the process  $\gamma~ g(k) \rightarrow q ~ \bar{q}$. Here
the gluon is off-shell, quasi transverse, with a virtuality 
$k^2 \, \simeq \kb^2$. One also has to introduce the unintegrated gluon
distribution density at a factorization scale $Q_0^2$, which is related to the usual gluon distribution by
\beq
\label{gluondist}
G(x_{bj},Q^2,Q_0^2) = \int^{Q^2}_{0} d^2 \kb \, {\cal F}(x_{bj},\kb,Q_0^2).
\eq
We first deal with a dipole of transverse size $x_{01}$, which can be part of an heavy onium (i.e. an heavy $q \bar q$ pair) or extracted from a proton as will be emphazied later. The $k_T$-factorization reads, for the total $\gamma^* - dipole$ cross-section,
\beq
\label{kt1}
Q^2 \sigma_{\gamma^*}^{d} (x_{bj},Q^2;x_{01}^2) = \int d^2\kb \int^{1}_{0}
\frac{d z}{z}\,  \hat{\sigma}_{\gamma g} (x_{bj}/z,\frac{\underline k^2}{Q^2})\, {\cal F}(z,\kb;x_{01}^2).
\eq
But we have now to couple the gluon to the softest dipole which arise in the cascade.
This can be fulfilled by using a second $k_T$-factorization.
It involves the elementary
Born  cross-section $\hat{\sigma}_{\gamma d}/k^2$ of the process  $d(\xb)~ g(k) \rightarrow  ~ d(\xb)$ for a dipole of transverse size $\xb$ and a soft gluon of virtuality $\kb^2$.
This $k_T$-factorization can be expressed by
\beq
\label{kt2}
\kb^2 {\cal F}(z,\kb;x_{01}^2) = \int \frac{d^2\xb}{\xb^2} \int^{1}_{0}
\frac{d z'}{z'} \, n (\xb_{01},\xb,z') \, \hat{\sigma}_{\gamma d} (z/z',\xb^2 \kb^2),
\eq
where $n (\xb_{01},\xb,z)$ is the distribution density of dipoles of transverse
size $\xb$ with the smallest light-cone momentum in the pair equal to $z p_+$ in a dipole of transverse
size $\xb_{01}$, of total momentum $p_+$. Computing $\hat{\sigma}_{\gamma d}$ with an eikonal coupling of the 
gluon \cite{nprw}, one gets 
\beqa
\label{kt1kt2}
Q^2 \sigma_{\gamma^*}^{d} (x_{bj},\frac{Q^2}{x_{01}^2}) = \int d^2\kb \int^{1}_{0}
\frac{d z}{z} \, \hat{\sigma}_{\gamma g} (x/z,\frac{\kb^2}{Q^2}) \int^{ }_{ } \frac{d^2\xb}{\xb^2} \, n (\xb_{01},\xb,z) \nonumber \\
\times 4 \, \pi \, \bar{\alpha} \, \frac{2 \, C_F}{(2 \pi)^4} \, (2 - e^{i \kb . \xb} -  e^{-i \kb . \xb}) \frac{1}{\kb^2}.
\eqa
Following \cite{mueller1}, we now introduce the double Mellin-transform of $n$ with respect to the rapidity
$Y = \ln z_1/z$ and the transverse size $\xb$ ($z_1 p_+$ is the light-cone momentum of the quark part of the dipole $(\xb_{01})$):
\beq
\label{mellinY}
n(\xb_{01},\xb,Y) = \int^{ }_{ } \frac{d \omega}{2 i \pi} e^{\omega Y} n_{\omega}(\xb_{01},\xb).
\eq
and
\beq
\label{mellinxb}
n_{\omega}(\xb_{01},\xb) = \int^{ }_{ } \frac{d \gamma}{2 i \pi} \left(\frac{x_{01}}{x} \right)^{2 \gamma} n_{\omega}(\gamma)
\eq
Note that (\ref{mellinxb}) is a decomposition into conformal three-points correlation functions,
due to the conformal invariance of the dipole integral kernel.  
One then gets \cite{mueller1}
\beq
\label{nnuomega}
n_{\omega}(\gamma) = \frac{2}{\omega - \frac{4 \bar{\alpha} C_F}{\pi} \chi(\gamma)}
\eq
where
\beq
\label{chi}
\chi(\gamma) = \Psi(1) - \de \Psi(\gamma) -  \de \Psi(1 - \gamma) \quad {\rm and} \quad
\Psi(\gamma) = \frac{d \ln \Gamma}{d \gamma}.
\eq
$\hat{\sigma}_{\gamma g}$ has been calculated for different polarizations of the incoming photon in \cite{catani}.
We introduce the corresponding double Mellin transform of this cross-section, namely
\beq
\label{hw}
4 \pi^2 \alpha_{e.m} h_{\omega}(\gamma) = \gamma \int^{\infty}_{0} \frac{d \kb^2}{\kb^2} \left(\frac{\kb^2}{Q^2}\right)^{\gamma} 
\hat{\sigma}_{\omega}(\frac{\kb^2}{Q^2})
\eq
or equivalently  
\beq 
\label{hinv}
\hat{\sigma}_{\omega}\left(\frac{l^2}{Q^2}\right)= 4 \pi^2 \alpha_{e.m} \int^{ }_{ } \frac{d \gamma}{2 i \pi}
 \left(\frac{l^2}{Q^2}\right)^{-\gamma} \frac{1}{\gamma} h_{\omega}(\gamma).
\eq
The expression for $\sigma_{\gamma^*}^{d}$ now reads
\beqa
\label{kt3}
&&\hspace{-1.5cm} Q^2 \sigma_{\gamma^*}^{d} (x_{bj},Q^2; x_{01}^2) = 4 \pi^2 \alpha_{e.m} \frac{\bar{\alpha} C_F}{2 \pi^3} \int^{ }_{ }  d^2\kb \int^{ }_{ } \frac{d \gamma'}{2 i \pi}  \int \frac{d \gamma}{2 i \pi} \int \frac{d \omega}{2 i \pi} e^{\omega Y} \left(\frac{x_{01}}{x}\right)^{2 \gamma} \nonumber \\
&&\times \frac{2}{\omega - \frac{4 \bar{\alpha} C_F}{\pi} \chi(\gamma)}  \frac{h_{\omega}(\gamma')}{\gamma'}  \int^{ }_{ } \frac{d^2\xb}{\xb^2}  \left(\frac{\underline k^2}{Q^2}\right)^{-\gamma'} (2 - e^{i \kb . \xb} -  e^{-i \kb . \xb}) \frac{1}{\kb^2} .
\eqa
The integration with respect to the polar angle of $\xb$ can easily be performed and
leads to a Bessel function. One has then to integrate over $x$, namely
\beq
\label{integrationx}
\int \frac{d x}{x} 4 \pi (1 - J_0(kx))\left (\frac{x}{x_{01}} \right)^{-2 \gamma} =
4 \pi (k x_{01})^{2 \gamma} \frac{2 ^{-1 -2 \gamma}}{\gamma} \frac{\Gamma(1- \gamma)}{\Gamma(1 + \gamma)} \equiv 4 \pi (k x_{01})^{2 \gamma} v(\gamma).
\eq
The integration over $k$ gives $\gamma = \gamma'$.
Since $n_{\nu \omega}$ (formula (\ref{nnuomega})) exhibits a pole at 
$\omega_p = \frac{\alpha_S N_c}{\pi} \chi(\gamma)$, the $\omega$ integral finally yields
\beq
\label{gonium1}
\frac{Q^2} {4 \pi^2 \alpha_{e.m}} \sigma_{\gamma^*}^{d} (x_{bj},Q^2;x_{01}^2) = \frac{2 \bar{\alpha} N_c}{\pi} \int \frac{d \gamma}{2 i \pi}  
 h_{\omega_p}(\gamma) \frac{v(\gamma)}{\gamma} (\xb_{01}^2Q^2)^{\gamma} 
e^{\frac{\bar{\alpha} N_c}{\pi} \chi(\gamma) \ln \frac{1}{x_{bj}}}.
\eq
In the following we neglect the dependence of $h_{\omega_p}(\gamma)$ on $\omega_p$ \cite{catani}.
The initial dipole state is supposed to be well localized in transverse space,  and its transverse size gives a {\it perturbative scale}.
We then get for the total cross-section $\gamma^* - dipole (\xb_{01})$ and for the related structure function:
\beq
\label{gonium2}
\frac{Q^2} {4 \pi^2 \alpha_{e.m}} \sigma_{\gamma^*}^{d} (x_{bj},Q^2;x_{01}^2) =  F_{\gamma}^{d}(x_{bj},Q^2;x_{01}^2)
= \frac{2\bar{\alpha} N_c}{\pi} \int^{ }_{ } \frac{d \gamma}{2 i \pi} (Q^2 x_{01}^2)^{\gamma} 
 h(\gamma) \frac{v(\gamma)}{\gamma}  
e^{\frac{\bar{\alpha} N_c}{\pi} \chi(\gamma) \ln\frac{1}{x_{bj}}}.
\eq
In order to deal with deep inelastic scattering on a proton, we will suppose that inside the proton
there exist configurations of dipole type revealed by the high-energy process ($u \overline{u}$ for example) which are well
transversally localised, with a typical transverse scale $x_{01}^2$. The distribution 
$w(\gamma,x_{01}^2;Q_0^2)$  of such
configurations cannot be computed perturbatively, but, using renormalisation group properties, this 
quantity is expected to have the following behaviour (see \cite{wil} for a similar analysis applied to the leading order QCD evolution equation):
\beq
\label{w}
w(\gamma,x_{01}^2;Q_0^2) = w(\gamma)(x_{01}^2 \, Q_0^2)^{-\gamma}.
\eq
$Q_0^2$ is a scale which is typical of the proton and expected to be non-perturbative. 
We get the final result:
\beqa
\label{crosshadronnonpert}
F^{prot}(x_{bj},Q^2;Q_0^2) = 2 \frac{\bar{\alpha} N_c}{\pi} \int^{ }_{ } \frac{d \gamma}{2 i \pi}  
 h(\gamma) \frac{v(\gamma)}{\gamma}  w(\gamma) \left(\frac{Q^2}{Q_0^2}\right)^{\gamma} 
\exp\left(\frac{\bar{\alpha} N_c}{\pi} \chi(\gamma) \ln\frac{1}{x_{bj}}\right).
\eqa
At this stage some comments are in order:
i)We insist on
the point that these initial pertubative configurations, even if they cannot be described easily, 
should in principle be present inside the proton, and their distribution non-negligible (in contrary 
to $c\overline{c}$ pairs for example, which, from $J/\Psi$ production, are known to be less relevant). Their evolution over a large range of rapidity is 
responsible for the cascade which contains the physics of the BFKL Pomeron. ii)This BFKL dynamics can
also be implemented when using the original BFKL kernel combined with Regge factorization. It involves
an initial unintegrated gluonic distribution, which is then evolved to lower values of $x_{bj}$ by the
cascade of reggeized gluons. Here again, this initial distribution cannot be predicted and has to be put by hand \cite{agkms}.

We now apply formula (\ref{crosshadronnonpert}) to peculiar structure functions, namely
\beq
\label{predgen}
\left(\begin{array}{c}
F_T \\ F_L \\ F_G \end{array} \right) 
= \frac{2\bar{\alpha} N_c}{\pi} \int^{ }_{ }  \frac{d \gamma}{2 i \pi}  
  \left(\frac{Q^2}{{Q_0}^2}\right) ^{\gamma} 
\exp\left(\displaystyle \frac{\bar{\alpha} N_c}{\pi} \chi(\gamma) \ln\frac{1}{x_{bj}}\right) \left(\begin{array}{c}
h_T \\ h_L \\ 1 \end{array} \right)
\frac{v(\gamma)}{\gamma} w(\gamma)
\eq
where $F_{T(L)}$ is the structure function corresponding to transverse (longitudinal) photons and $F_G$ the gluon structure function.
The coefficient functions 
\beq
\label{defh}
\left(\begin{array}{c}
h_T \\ h_L \end{array} \right) = \frac{\bar{\alpha}}{ 3 \pi \gamma} 
\frac{(\Gamma(1 - \gamma) \Gamma(1 + \gamma))^3}{\Gamma(2 - 2\gamma) \Gamma(2 + 2\gamma)} \frac{1}{1 - \frac{2}{3} \gamma} \left( \begin{array}{c} (1 + \gamma)
(1 - \frac{\gamma}{2}) \\ \gamma(1 - \gamma) \end{array} \right)
\eq
were computed in ref \cite{catani}. 
The coupling constant $\bar{\alpha}$ enters the coupling of the virtual photon to the off-shell gluon, and thus in formula
(\ref{defh}). It also enters the effective coupling of the dipole kernel. We shall take it identical in both cases. We also keep $\bar{\alpha}$ non running, even if it might be possible to implement the running of $\bar{\alpha}$ in the dipole cascade \cite{ross}. Indeed, this would involve next-to-leading terms which are beyond this model and not already known.
Nethertheless, the running of $\bar{\alpha}$ is not expected to drastically change the shape of the dipole cascade.

\noindent
When $x_{bj}$ is small, the $\gamma-$integration can be performed by the
 steepest-descent method. Considering $w(\gamma)$
as smooth and regular near $\gamma = \de$, we expand the $\chi$ function around the asymptotic anomalous dimension $\gamma = \de$.
We obtain a saddle point at  
\beqa
\label{saddle}
\gamma_s &=& \frac{1}{2} \left( 1 - a \ln \frac{Q}{Q_0}\right), \quad
\mbox {where} \quad  a  = \left(\frac{\bar{\alpha} N_c}{\pi} 7 \zeta(3) \ln\frac{1}{x_{bj}}\right) ^{-1}.
\eqa
The approximation of expanding $\chi(\gamma)$ around $\de$ is valid when 
\beq
\label{applicabilite}
a \ln\left(\frac{Q}{Q_0}\right) \simeq \ln \frac{Q}{Q_0}/\ln\frac{1}{x_{bj}} \ll 1,
\eq
that is the small $x_{bj}$, moderate $Q/Q_0$ kinematical domain.
This yields
\beq
\label{predF2}
F_2 \equiv F_T + F_L = C a^{1/2} \frac{Q}{Q_0} \exp \left( (\alpha_{P} -1) \ln\frac{1}{x_{bj}} - \frac{a}{2} \ln^2 \frac{Q}{Q_0} \right),
\eq
where
\beq
\label{alphap}
\alpha_{P} -1 = \frac{4 \bar{\alpha} N_{C} \ln 2}{\pi}
\eq
Thus, $F_2$ depends only on 3 parameters, $C, \ Q_0 \  {\rm and} \  \alpha_{P}.$
Fitting $F_2$ with this form, one can get a prediction for $F_G$ and $R = F_L / F_T$. Namely,
\beq
\label{fgf2}
\frac{F_G}{F_2} = \left. \frac1{h_T + h_L}\right|_{\gamma = \gamma_s}\equiv \frac{3 \pi \gamma_s}{\bar{\alpha}} \frac{1 - \frac{2}{3}\gamma_s}{1 + \frac{3 }{2}\gamma_s - \frac{3}{2}\gamma_s^2} \frac{\Gamma(2 - 2 \gamma_s) \Gamma(2 + 2 \gamma_s)}{(\Gamma(1 - \gamma_s)\Gamma(1 + \gamma_s))^3}  
\eq
and
\beq
\label{R}
R = \frac{h_L}{h_T}(\gamma_s) = \frac{\gamma_s (1 - \gamma_s)}{(1 + \gamma_s)(1 - \frac{\gamma_s}{2})},
\eq
where  $\gamma_s$ is given  by the expression (\ref{saddle}).
Note that the overall non-perturbative normalization $C$ does not enter $R$ and $F_G/F_2$.

\section{$F_2$ fit and prediction for $F_G$ and $R$}

\begin{figure}[htbp]
\begin{picture}(500,470)(0,0)
\centering
\put(0,-10){\epsfysize=20.0cm{\hspace{1.8CM}\centerline{\epsfbox{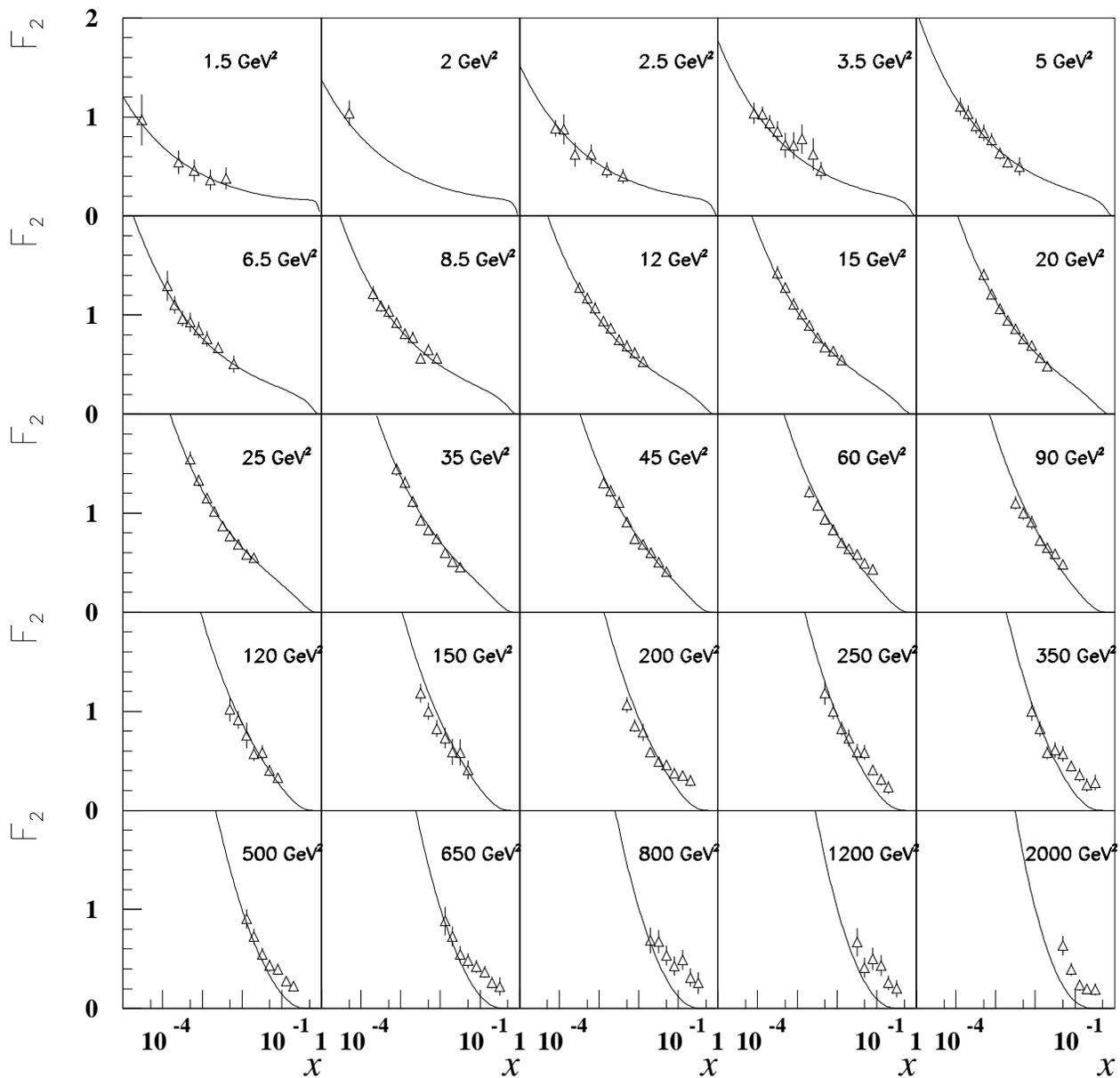}}}}
\end{picture}
\caption{Comparison of the fit of the H1 data for $Q^2 \leq 150 \, GeV^2$ with all the 1994 H1 data. The discrepancy at high $x_{bj}$ and high $Q^2$ is outside the validity of our model.}
\label{F2}
\end{figure}

A fit of the 1994 H1  data \cite{h1} has been performed. We have kept only bins in $Q^2$ with 
$Q^{2} \leq 150$ $GeV^{2}$, in order to stay in the range where the QCD dipole model is relevant and to insure that our saddle point method can be trusted (see the limit (\ref{applicabilite})). This point is discussed in ref. \cite{thesewal}. Note that our fit does
not depend strongly on that cut value. 
The fit parameters are $C$, $Q_0$ and $\alpha_P$. The obtained fit for $F_2$ is displayed in figure \ref{F2} (the highest $Q^2$ point at $5000 \, GeV^2$ is not displayed for convenience).
We get a $\chi^2$ of $101$ for $130$ points, and the following fitted parameters:
\begin{eqnarray}
&~& \alpha_{P}=1.282  \nonumber \\
&~& Q_{0}=0.627 \, GeV  \nonumber \\
&~& C=0.077.
\end{eqnarray}
The obtained value for the hard Pomeron intercept $\alpha_P$ is in agreement with other determinations applying BFKL dynamics to $F_2$ data at HERA \cite{agkms,h1}. The corresponding effective coupling constant is $\bar{\alpha} \simeq
.11$, close to $\alpha(M_{Z_0})$ used in the H1 QCD study. However, the obtained value for the Pomeron intercept is relatively smaller than 
the value one would expect ($\alpha_P \simeq 1.5$), and the
corresponding effective coupling is surprisingly small. This might be due to next-to-leading order
terms, responsible for this effective decrease of $\alpha_P$. This is indeed the case for example when one attempt to sum up both soft and collinear singularities and if one takes into account energy-momentum conservation
 \cite{unif,ekl}.  
The value of $Q_0$ corresponds to a transverse size of $.3$ fm, which is, as expected, a non-perturbative scale typical of deep inelastic scattering on a proton target. The non predictable parameter $C$ is linked to the non-perturbative input $w(\de)$.
It is quite remarkable that one can well  reproduce the data in the low-$Q^2$ range with only $3$ parameters. It confirms a previous study \cite{npr} using H1 and ZEUS 1993 data \cite{H1ZEUS}. Deviations from the fit at higher-$Q^2$ values have several origins. 
\begin{figure}[h!]
\begin{picture}(500,410)(0,0)
\centering
\put(0,-10){\epsfysize=17.0cm{\centerline{\hspace{3.5CM}\epsfbox{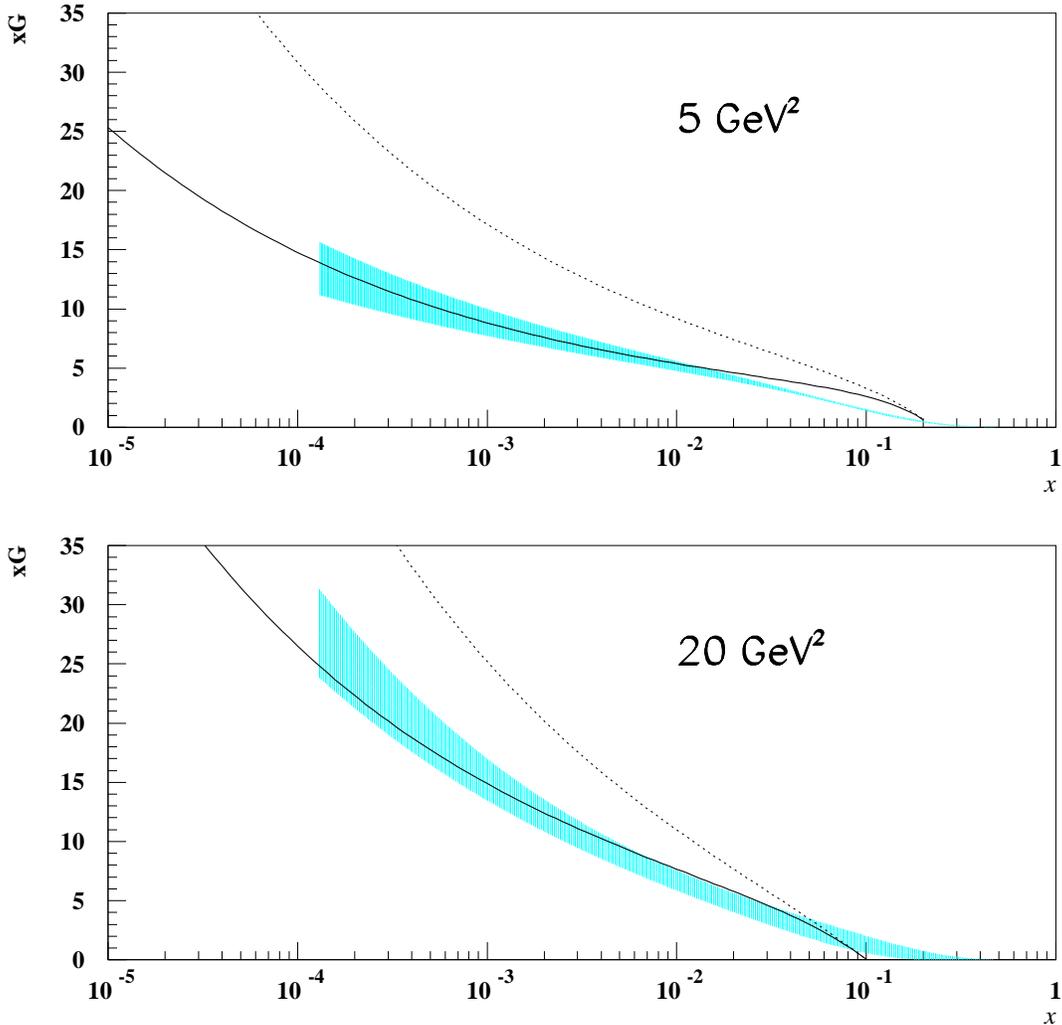}}}}
\end{picture}
\caption{Prediction for the gluon density $F_G(x_{bj},Q^2)$ (full line) compared with the H1 prediction based on their next-to-leading DGLAP QCD fit. In dotted line the one loop approximation is displayed.}
\label{GLUONS}
\end{figure}
 First, the valence contribution is not contained in the dipole model, and  it is well known that it should 
dominate at higher-$x_{bj}$ values. Because of the HERA kinematical constraints, this corresponds to bins at higher-$Q^2$ values.
Second, we used an expansion of the dipole eigenvalue around $\de$, which is valid only for moderate-$Q^2$ values. A phenomenological input for the valence contribution and a numerical treatment of the 
$\gamma$  integral is expected to improve the fit of the HERA data in the higher $Q^2$ range.

Relation (\ref{fgf2}) provides now a precise determination of the gluon density. We use for $Q_0$ and $\bar{\alpha}$ 
the values obtained from the $F_2$ fit. In figure \ref{GLUONS} we show the comparison between the published H1 determination of the gluon density (grey band), based on a next-to-leading order DGLAP analysis, and our prediction (full line).
\begin{figure}[htbp]
\begin{picture}(500,470)(0,0)
\centering
\put(10,-70){\epsfysize=21.0cm{\centerline{\hspace{1CM}\epsfbox{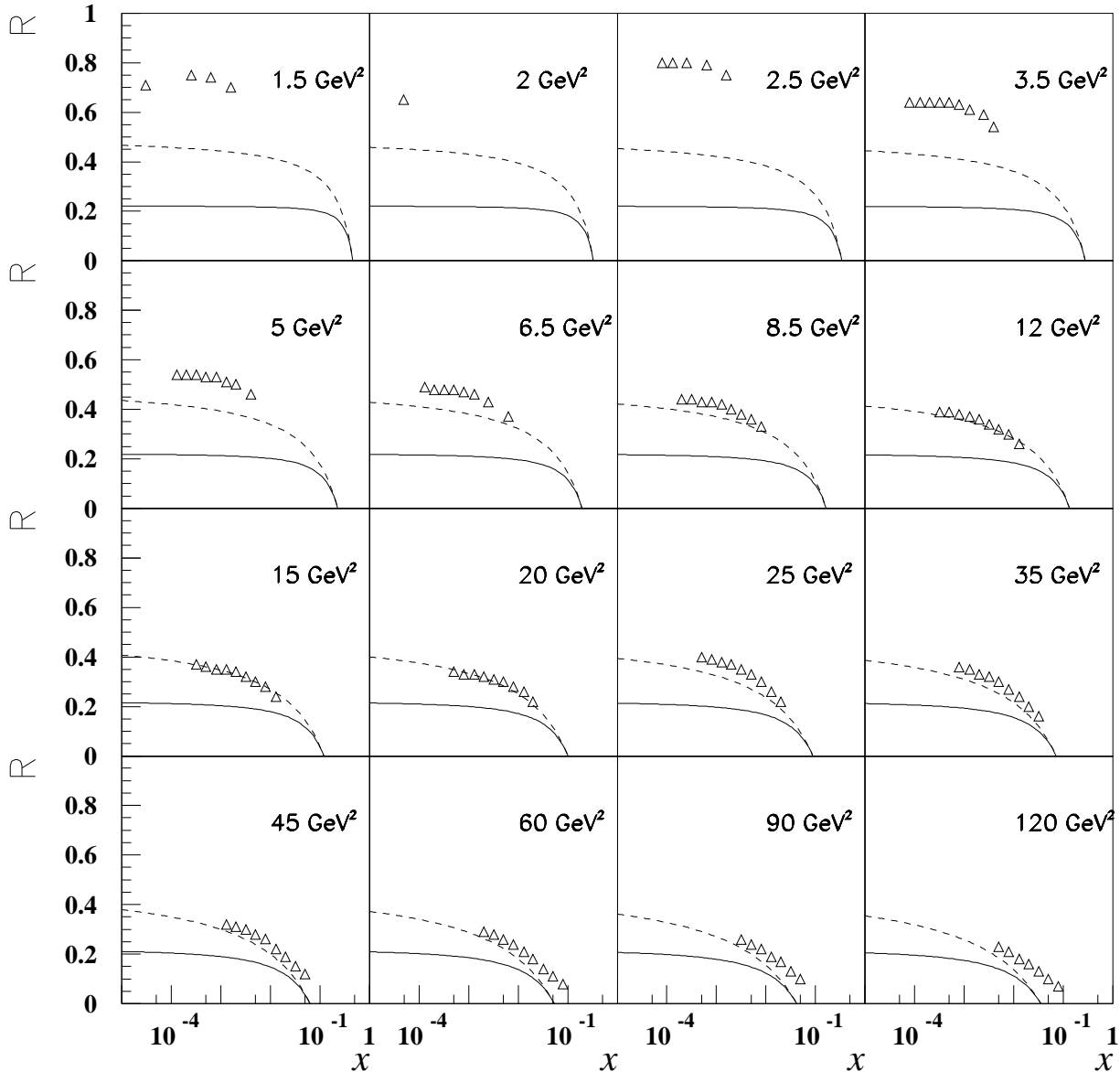}}}}
\end{picture}
\caption{Prediction for the $R$ (full line) compared to the Altarelli-Martinelli prediction used in the H1 study. In dotted line the one loop approximation is displayed.}
\label{Rcompare}
\end{figure}
We also display the prediction obtained from first order perturbative expansion of the 
coefficient functions $h$, in order to see the resummation effects. We notice that for $x \leq 10^{-2}$, our prediction is in the range quoted by H1 collaboration \cite{h1}. Thus, the gluon distribution
does not provide a good tool in distinguishing between DGLAP and BFKL dynamics.

The prediction for $R$ is rather different. In figure \ref{Rcompare}, we display the Altarelli-Martinelli \cite{am} prediction for $R$,  used in the H1 study \cite{h1},
in comparison with our prediction. 
We also show up the resummation effect for $h$.
The discrepancy between AM and BFKL predictions is rather strong. In particular, the limiting value of the dipole model is $2/9$, that is much lower than the AM prediction \cite{ad}. $R$ could thus provide a good observable in distinguishing between the
two possible evolutions.

\section{Conclusion}

In this contribution we have shown that combining the dipole model with $k_T$-factorization,
it is possible to implement the BFKL dynamic in order to describe the proton structure functions.
With only 3 parameters, the most recent 1994 H1 data can be well described in the range $Q^2 \leq 150 \ GeV^2$. We obtain a prediction for $F_G$ which is close to the one based on next-to-leading order DGLAP equations. In contrast, the prediction for $R$ is significantly different. Measurement
of this quantity would thus be of great interest in the low-$x_{bj}$ range.  
\\

\noindent
{\bf Aknowledgments}

\noindent
The most part of what is written in the present contribution comes from a fruitful collaboration with
Henri Navelet, Robi Peschanski and Christophe Royon. I thank Jan Kwieci\' nski for discussion  during the conference. I also thank Al~Mueller for comments.
\vskip.5cm

\noindent{\large {\bf References}}

\end{document}